\documentclass[twoside,twocolumn,9pt]{article}
\usepackage{extsizes}
\usepackage[super,sort&compress,comma]{natbib} 
\usepackage[version=3]{mhchem}
\usepackage[left=1.5cm, right=1.5cm, top=1.785cm, bottom=2.0cm]{geometry}
\usepackage{balance}
\usepackage{mathptmx}
\usepackage{sectsty}
\usepackage{graphicx} 
\usepackage{lastpage}
\usepackage[format=plain,justification=justified,singlelinecheck=false,font={stretch=1.125,small,sf},labelfont=bf,labelsep=space]{caption}
\usepackage{float}
\usepackage{fancyhdr}
\usepackage{fnpos}
\usepackage[english]{babel}
\usepackage[utf8]{inputenc}
\addto{\captionsenglish}{%
  
}
\usepackage{array}
\usepackage{droidsans}
\usepackage{charter}
\usepackage[T1]{fontenc}
\usepackage[usenames,dvipsnames]{xcolor}
\usepackage{setspace}
\usepackage[compact]{titlesec}
\usepackage{hyperref}

\usepackage{epstopdf}

\usepackage{amsmath}
\usepackage{mathtools}
\usepackage{float}
\usepackage{braket}
\usepackage{bm}
\usepackage{siunitx} 
\usepackage{boldline} 

\usepackage{multicol, makecell, booktabs, array}

\DeclareSIUnit \angstrom {\text{Å}}
\newcommand{\ag}{\si{\angstrom}}

\definecolor{cream}{RGB}{222,217,201}

\begin{document}

\pagestyle{fancy}
\thispagestyle{plain}
\fancypagestyle{plain}{
\renewcommand{\headrulewidth}{0pt}
}

\makeFNbottom
\makeatletter
\renewcommand\LARGE{\@setfontsize\LARGE{15pt}{17}}
\renewcommand\Large{\@setfontsize\Large{12pt}{14}}
\renewcommand\large{\@setfontsize\large{10pt}{12}}
\renewcommand\footnotesize{\@setfontsize\footnotesize{7pt}{10}}
\makeatother

\renewcommand{\thefootnote}{\fnsymbol{footnote}}
\renewcommand\footnoterule{\vspace*{1pt}%
\color{cream}\hrule width 3.5in height 0.4pt \color{black}\vspace*{5pt}} 
\setcounter{secnumdepth}{5}

\makeatletter 
\renewcommand\@biblabel[1]{#1}            
\renewcommand\@makefntext[1]%
{\noindent\makebox[0pt][r]{\@thefnmark\,}#1}
\makeatother 
\renewcommand{\figurename}{\small{Fig.}~}
\sectionfont{\sffamily\Large}
\subsectionfont{\normalsize}
\subsubsectionfont{\bf}
\setstretch{1.125} 
\setlength{\skip\footins}{0.8cm}
\setlength{\footnotesep}{0.25cm}
\setlength{\jot}{10pt}
\titlespacing*{\section}{0pt}{4pt}{4pt}
\titlespacing*{\subsection}{0pt}{15pt}{1pt}

\fancyfoot{}
\fancyfoot[LO,RE]{\vspace{-7.1pt}\includegraphics[height=9pt]{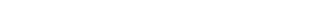}}
\fancyfoot[CO]{\vspace{-7.1pt}\hspace{13.2cm}\includegraphics{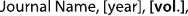}}
\fancyfoot[CE]{\vspace{-7.2pt}\hspace{-14.2cm}\includegraphics{head_foot/RF}}
\fancyfoot[RO]{\footnotesize{\sffamily{1--\pageref{LastPage} ~\textbar  \hspace{2pt}\thepage}}}
\fancyfoot[LE]{\footnotesize{\sffamily{\thepage~\textbar\hspace{3.45cm} 1--\pageref{LastPage}}}}
\fancyhead{}
\renewcommand{\headrulewidth}{0pt} 
\renewcommand{\footrulewidth}{0pt}
\setlength{\arrayrulewidth}{1pt}
\setlength{\columnsep}{6.5mm}
\setlength\bibsep{1pt}

\makeatletter 
\newlength{\figrulesep} 
\setlength{\figrulesep}{0.5\textfloatsep} 

\newcommand{\topfigrule}{\vspace*{-1pt}%
\noindent{\color{cream}\rule[-\figrulesep]{\columnwidth}{1.5pt}} }

\newcommand{\botfigrule}{\vspace*{-2pt}%
\noindent{\color{cream}\rule[\figrulesep]{\columnwidth}{1.5pt}} }

\newcommand{\dblfigrule}{\vspace*{-1pt}%
\noindent{\color{cream}\rule[-\figrulesep]{\textwidth}{1.5pt}} }

\makeatother

\twocolumn[
  \begin{@twocolumnfalse}
{\includegraphics[height=30pt]{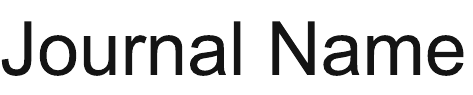}\hfill\raisebox{0pt}[0pt][0pt]{\includegraphics[height=55pt]{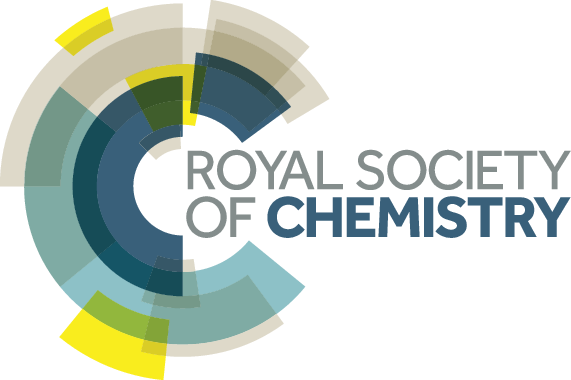}}\\[1ex]
\includegraphics[width=18.5cm]{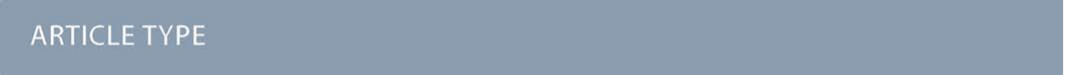}}\par
\vspace{1em}
\sffamily
\begin{tabular}{m{4.5cm} p{13.5cm} }

\includegraphics{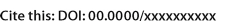} & \noindent\LARGE{\textbf{Mn$_2$C MXene Functionalized by Oxygen is a Semiconducting Antiferromagnet and Efficient Visible Light Absorber$^\dag$}} \\
\vspace{0.3cm} & \vspace{0.3cm} \\

 & \noindent\large{Jiří Kalmár\textit{$^{a}$} and František Karlický$^{\ast}$\textit{$^{a}$}} \\

\includegraphics{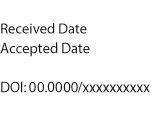} & \noindent\normalsize{
Manganese-based MXenes are promising two-dimensional materials due to the broad palette of their magnetic phases and the possibility of experimental preparation because the corresponding MAX phase was already prepared. 
Here, we systematically investigated geometrical conformers and spin solutions of oxygen-terminated \ce{Mn2C} MXene and performed subsequent many-body calculations to obtain reliable electronic and optical properties. 
Allowing energy-lowering using the correct spin ordering via supercell magnetic motifs is essential for the \ce{Mn2CO2} system. 
The stable ground-state \ce{Mn2CO2} conformation is antiferromagnetic (AFM) one with zigzag lines of up and down spins on Mn atoms. The AFM nature is consistent with the parent MAX phase and even the clean depleted \ce{Mn2C} sheet. 
Other magnetic states and geometrical conformations are energetically very close, providing state-switching possibilities in the material. 
Subsequent many-body GW and Bethe-Salpeter equation (BSE) calculations provide indirect semiconductor characteristics of AFM \ce{Mn2CO2} with a fundamental gap of 2.1~eV (and a direct gap of 2.4~eV), the first bright optical transition at 1.3~eV and extremely strongly bounded (1.1~eV) first bright exciton. 
\ce{Mn2CO2} absorbs efficiently the whole visible light range and near ultraviolet range (between 10 - 20\%).
} \\

\end{tabular}

 \end{@twocolumnfalse} \vspace{0.6cm}

  ]

\renewcommand*\rmdefault{bch}\normalfont\upshape
\rmfamily
\section*{}
\vspace{-1cm}


\footnotetext{\textit{$^{a}$~~Department of Physics, Faculty of Science, University of Ostrava, 30. dubna 22, 7013 Ostrava, Czech Republic. Tel: +420 553 46 2155; E-mail: frantisek.karlicky@osu.cz}}

\footnotetext{\dag~Electronic Supplementary Information (ESI) available: [details of any supplementary information available should be included here]. See DOI: 00.0000/00000000.}




\section{Introduction}
MXenes, a family of recent two-dimensional (2D) carbides, nitrides, and carbonitrides, embody various interesting properties promising for technical applications.  
Besides more than 20 MXenes prepared experimentally, theoretical modeling suggested tens of others and such a tool became necessary in the design of new MXenes and their properties.\cite{Anasori2020Book,Hantanasirisakul2018,Anasori2022,Lim2022}
The typical composition of MXenes is M$_n$X$_{n-1}$T$_x$ with $1\le n \le4$ and $x \le 2$, where M is metal, X is carbon or nitrogen, and T is the surface termination which includes groups 16 and 17 of the periodic table or hydroxyl and imido groups.
2D MXene's properties are sensitive to composition, termination, or external conditions and provide a rich set of its phases: metals, semiconductors,\cite{Ketolainen2022} ferromagnets,\cite{Si2015, Zhang2017, Tan2021, Venosova2024} antiferromagnets,\cite{Venosova2024, Lv2020, Gorkan2022, GarcaRomeral2023, Sakhraoui2022} topologic insulators\cite{Champagne2021} or excitonic insulators\cite{Kumar2023} with the possibility of magnetization modulation by electric field\cite{Xin2023, Xin2023-2} or lattice defects.\cite{Lv2020}
Here we focus on less explored manganese-based MXenes as they are extremely interesting for the energetic competition of magnetic states. 
This MXene subfamily is still rather a hypothetical one, however, it is proposed that Mn$_2$CT$_x$ MXenes can be prepared experimentally from the already existing parent Mn$_2$GaC MAX phase\cite{Ingason2014, Dahlqvist2016, Novoselova2018, Thorsteinsson2023} by exfoliation techniques. 
Also, other Mn-containing magnetic MAX phases as (Cr,Mn)$_2$AlC or (Mo,Mn)$_2$GaC were prepared.\cite{Mockute2014, Salikhov2017}
The Mn-based MXenes are typically predicted as conductors\cite{Champagne2021, Kozak2023} and the standard prediction tool for materials modeling, the density functional theory (DFT), is robust and generally accepted. 
However, DFT can sometimes predict different or opposite properties when using its various levels of implementation (mainly various density functionals) for complicated systems (e.g. transition metals containing ones). 
Therefore, when recent works proved, that at least one Mn-based MXene, namely oxygen-terminated Mn$_2$C MXene, is a semiconductor,\cite{Ketolainen2022, Gao2022, Chen2021, Zhou2017} the need for a deeper understanding of the systems was growing. 
In addition, the standard generalized gradient approximation (GGA) of DFT showed antiferromagnetic (AFM) behavior of \ce{Mn2CO2}, while recent (higher level) hybrid DFT calculation predicted the ferromagnetic (FM) phase.\cite{Ketolainen2022} 
Moreover, if magnetism is properly treated, magnetic motifs larger than a unit cell should be considered too.\cite{Kolos2020}
Finally, we showed recently the unusually high absorptance in \ce{Mn2CO2} of $A\approx 10-20 \%$ in the visible 1-3~eV range\cite{Ketolainen2022} (however, in simple unit cell using time-dependent DFT), so the material would be interesting also technologically.

The uncertainties mentioned, the complicated magnetism, the existence of geometric and magnetic conformers, and the close energetic levels of the corresponding various electronic states should be reconciled by systematical study and more accurate methods (including many-body methods). 
Especially, if the material is expected to be an excellent absorber and possibly an easy switch between FM and AFM phases, both technologically promising for device applications.
We therefore used state-of-the-art methods to discover the right nature of \ce{Mn2CO2} and reliably determine its promising properties. 
Many-body perturbation theory similes experiments like ARPES and EELS in that the former precisely captures the same single-particle spectral function. 
Within this framework in GW approximation,\cite{Hedin1965} the light absorption can be fully understood through the solution of the excitonic equation of motion, the Bethe-Salpeter equation (BSE).\cite{Bethe1951} 
Firstly, we systematically investigated many different spin solutions and geometrical conformers of \ce{Mn2CO2}, determined the ground state, and confirmed its stability by phonon analysis and molecular dynamics. 
Then, we used well-converged advanced many-body methods to determine the fundamental gap, optical gap, and exciton binding energy. 
Finally, we determined the optical properties and classified \ce{Mn2CO2} as an efficient absorber of the visible and near ultraviolet (UV) part of the sun spectrum. 

\section{Computational methods}
\label{methods}
All calculations have been performed using the periodic density functional theory code Vienna \textit{Ab initio} Simulation Package (VASP)\cite{kresse_ab_1993, kresse_efficiency_1996,kresse_efficient_1996,kresse_norm-conserving_1994} in versions 6.2.1 and 6.3.0. 
The spin-polarized DFT Kohn-Sham equations have been solved variationally in a plane-wave basis set using the projector-augmented-wave (PAW) method.\cite{Blochl1994} 
For structural optimization, ground-state calculations, and band structure calculations the Perdew-Burke-Ernzerhof (PBE) density functional\cite{PBE} in generalized gradient approximation (GGA) was used as well as meta-GGA Strongly Constrained and Appropriately Normed (SCAN)\cite{SCAN} density functional. 
Later, an advanced hybrid density functional HSE06 was used to further compare energetically close solutions.
The convergence criterion for the electronic self-consistency cycle was in all cases set to $10^{-7}$~eV/cell and the structural optimization converged within $10^{-2}$~eV/$\ag$. 
The plane-wave cutoff energy was set to 500 eV for all calculations uniformly. 
GW sets of PAWs were used in all calculations and only valence electrons were considered for C and O atoms (2$s^2$1$p^2$ and 2$s^2$1$p^4$, respectively) but the semi-core $s$ and
$p$ states were added for Mn atoms (3$s^2$3$p^6$3$d^5$4$s^2$).\cite{rem_paw} 
For a standard unit cell, $12\times12\times1$ and $24\times24\times1$ k-point grids were used for relaxation and ground-state calculation, respectively. 
For subsequent magnetic $2\times2$ supercells, the grid was changed to $6\times6\times1$ and $12\times12\times1$, respectively, to maintain a constant k-point grid density. 
In all band structure calculations, the standard k-point path of $\Gamma$-M-K-$\Gamma$ for 2D hexagonal systems was used. 
\ce{Mn2CO2} supercells and its phonon dispersion spectra were generated using the \texttt{Phonopy} code.\cite{Togo2015} 
The Hessian matrix for phonon calculation was obtained using density functional perturbation theory (DFPT). 
\textit{Ab initio} molecular dynamics was performed on $6\times 6\times 1$ supercells at 400~K, using the Andersen thermostat and time steps of 2~fs. 

The quasi-particle energies $\epsilon_{n\bm{k}}^{\mathrm{GW}}$ were calculated as first-order corrections to the Kohn-Sham energies $\epsilon_{n\bm{k}}$ (i.e., using so-called single-shot G$_0$W$_0$ variant).\cite{Shishkin2006} 
The quasiparticle gap was computed as $\Delta^\mathrm{GW}=\epsilon_\mathrm{CBM}^{\mathrm{GW}}-\epsilon_\mathrm{VBM}^{\mathrm{GW}}$, where CBM stands for conduction band minimum, and VBM denotes valence band maximum. 
We note that G$_0$W$_0$ fundamental gaps obtained on top of GGA PBE densities (G$_0$W$_0$@PBE) are very accurate (when perfectly converged) for 2D materials composed of $sp$ elements, as we proved recently by comparison with experiment\cite{Kolos2019, Kolos2022} or by direct comparison with the independent stochastic many-body fixed-node diffusion Monte Carlo (FNDMC) method.\cite{Dubecky2020}
Recently, G$_0$W$_0$@PBE gap was in agreement with FNDMC also in the case of nonmagnetic direct semiconducting scandium-based carbide (MXene).\cite{Dubecky2023} 
Here, we are treating a complicated 3$d$-metal-element-containing antiferromagnetic/ferromagnetic system, so the GGA PBE density for input wavefunction is not the best solution. 
To account for the energy of localized 3d orbitals of transition metal (TM) atoms properly, the Hubbard “U” correction is typically employed, and it is recommended for \ce{Mn2CO2} too.\cite{He2016}
To avoid the empirical choice of U value, we selected more general meta-GGA SCAN functional, as both SCAN and PBE+U provide similar magnetic moments and bandgaps for 3$d$-metal ferroelectrics and multiferroics\cite{Sun20216} (see also section \ref{dft-gs}). 
In opposite, we did not build G$_0$W$_0$ on top of hybrid functionals as HSE06, mixing some fraction of exact Hartree-Fock exchange, because HSE06 overestimate \ce{Mn2CO2} magnetic moments\cite{Ketolainen2022} and can fail because of multireference system\cite{Cho2023} or due to tradeoffs between over-delocalization and under-binding.\cite{Janesko2021} 
In addition, subsequent G$_0$W$_0$@HSE06 gaps often overestimate experiment, e.g., for antiferromagnetic hematite $\alpha$-\ce{Fe2O3}.\cite{Liao2011} 
G$_0$W$_0$@HSE06 approach is typically used just in the case of small-gap semiconductors, where PBE provides an artificially negative gap preventing subsequent use of perturbative G$_0$W$_0$@PBE.\cite{Grumet2018, Kalmar2024}

The BSE was used in eigenvalue problem form\cite{Albrecht1998} for insulating materials with occupied valence bands ($v$ index), and completely unoccupied conduction bands ($c$),
\begin{equation}
\label{BSE}
\begin{multlined}
(\epsilon_{c\bm{k}}^{\mathrm{GW}}-\epsilon_{v\bm{k}}^{\mathrm{GW}})A_{cv \bm{k}}^\lambda
+\sum_{c'v'\bm{k}'}[2\langle \phi_{c\bm{k}}\phi_{v\bm{k}} \lvert \nu \rvert \phi_{c'\bm{k}'}\phi_{v'\bm{k}'}\rangle \\ 
-\langle \phi_{c\bm{k}}\phi_{c'\bm{k}'} \lvert W \rvert \phi_{v\bm{k}}\phi_{v'\bm{k}'}\rangle
]A_{c'v'\bm{k'}}^\lambda
=
E_{\mathrm{exc}}^\lambda A_{cv\bm{k}}^\lambda,
\end{multlined}
\end{equation}
where $\nu$ is the Coulomb kernel, $1/\lvert \bm{r}-\bm{r}'\rvert$, the eigenvectors $A_{cv \bm{k}}^\lambda$ correspond to the amplitudes of free electron-hole pair configurations composed of electron states $\rvert \phi_{c\bm{k}}\rangle$ and hole states $\rvert \phi_{v\bm{k}}\rangle$. 
I.e., the excitonic wave function stands as
\begin{equation}\label{exciton_wf}
\sum_{cv\mathbf{k}}A_{cv\mathbf{k}}^{\lambda}\ket{\phi_{c\mathbf{k}}}\ket{\phi_{v\mathbf{k}}}.
\end{equation}
The eigenenergies $E_{\mathrm{exc}}^\lambda$ correspond to the excitation energies (with optical gap $\Delta_\mathrm{opt}^\mathrm{BSE} \equiv E_{\mathrm{exc}}^{\lambda}$, for $\lambda$ from first nonzero transition, i.e., first bright exciton). 
The difference $E_\mathrm{b}=\Delta^{\mathrm{GW,dir}}-\Delta^{\mathrm{BSE}}_{\mathrm{opt}}$ is called exciton binding energy, where "dir" index denotes direct quasiparticle gap.

\section{Results and discussion}\label{results}
\subsection{Geometrical structures and magnetic solutions}\label{structure}

\begin{figure}[h]
\centering
  \includegraphics[width=7cm]{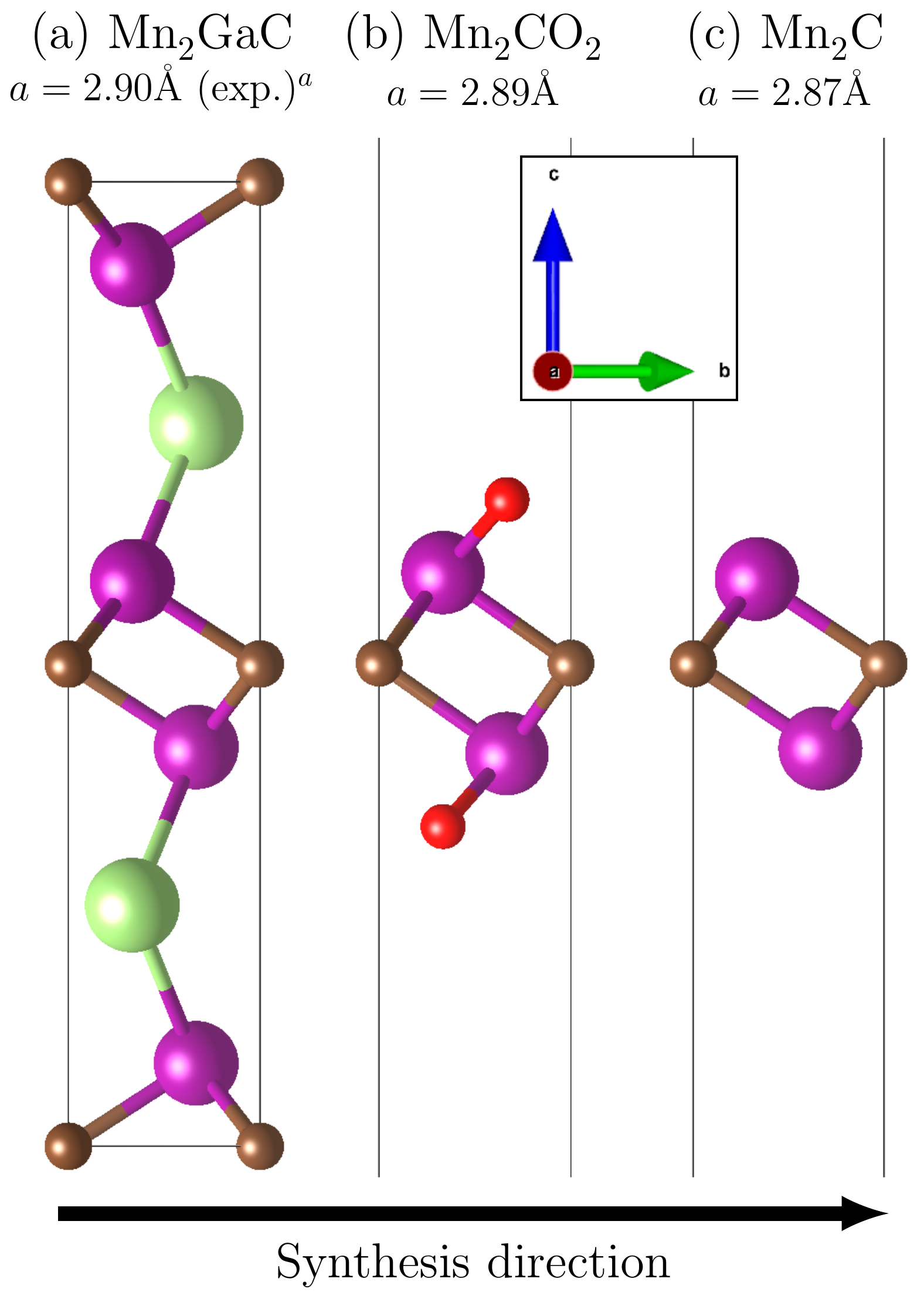}
  \caption{(a) \ce{Mn2GaC} precursor MAX phase, (b) functionalized \ce{Mn2CO2} MXene, and (c) a clean \ce{Mn2C} sheet, all with highlighted unit cell boundaries. Purple atoms = Mn, brown = C, red = O, light green = Ga. The black arrow signifies the typical progression of MXene's preparation stages. \\
  $^a$Experimental results measured at room temperature and at 150 K.\cite{Ingason2014, Dahlqvist2016}}
  \label{phases}
\end{figure}

The already prepared precursor MAX phase \ce{Mn2GaC} was measured to embody AFM behavior up to 507 K (Néel temperature),\cite{Novoselova2018} local magnetic moments of $\sim$1.7$\mu_B$ per Mn atom,\cite{Dahlqvist2016} and have a lattice constant $a = 2.90 \ \ag$ both at 150 K\cite{Dahlqvist2016} and room temperature.\cite{Ingason2014} 
During the etching processes, functional groups (-O, -F, -OH,\dots) cover both sides of the clean MXene layers (realistically with some percentage of vacancies). 
It can be expected that the adsorption of termination groups can slightly alter the lattice constant of the parent MAX phase. 
It is then possible to obtain MXene sheets with depleted surface terminations. Recently, Persson et al.\cite{Persson2018} have subjected the \ce{Ti3C2T}$_x$ MXene to vacuum thermal treatment with subsequent exposure to \ce{H2} gas and managed almost completely to deplete the surface of F-terminations. 
Oxygen terminations were removed only partly with the Ti:O ratio being 3:0.6 (i.e., 30\% surface coverage). 
Further depletion of terminations may be possible at higher \ce{H2} pressures than those accessible in the environmental transmission electron microscope (ETEM) used. 
All three mentioned stages of preparation can be seen in Figure \ref{phases} (experimental lattice constant compared to our calculated ones for oxygen-terminated and clean MXene). 
For reasons stated above, every conformation or spin solution of \ce{Mn2CO2} discussed further has a fixed lattice constant, which belongs to the ground-state \ce{Mn2CO2} MXene. 

\begin{figure}[h]
\centering
  \includegraphics[height=8cm]{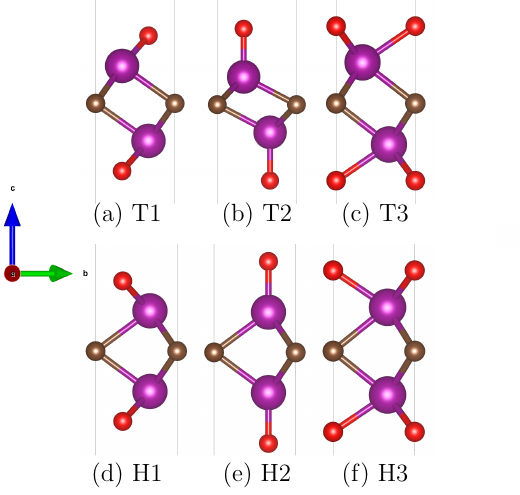}
  \caption{Trigonal (T) and hexagonal (H) geometric conformations of \ce{Mn2CO2} with varying positions (1--3) of oxygen atoms. Purple atoms = Mn, brown = C, red = O.}
  \label{geom_isomers}
\end{figure}

In our investigation, we have created 6 different geometric conformers of Mn$_2$CO$_2$ MXene (Figure \ref{geom_isomers}). 
These conformers correspond to the T- (trigonal) and H- (hexagonal) structures of transition metal chalcogenides and the numbers denote the positions of functional groups (oxygen in our case): hollow site (1), metal site (2), and carbon site (3). 
After conducting PBE relaxation on those conformers, we have found that H1 one in its AFM state is energetically the most stable. 
T2, T3, H2, and H3 conformers were all significantly higher in energy, but the energy difference between T1-AFM and H1-AFM conformers was only 0.06 eV. 
We have therefore concluded that a closer look at those geometries and possible magnetic states is needed.

\begin{figure}[h]
\centering
  \includegraphics[height=10cm]{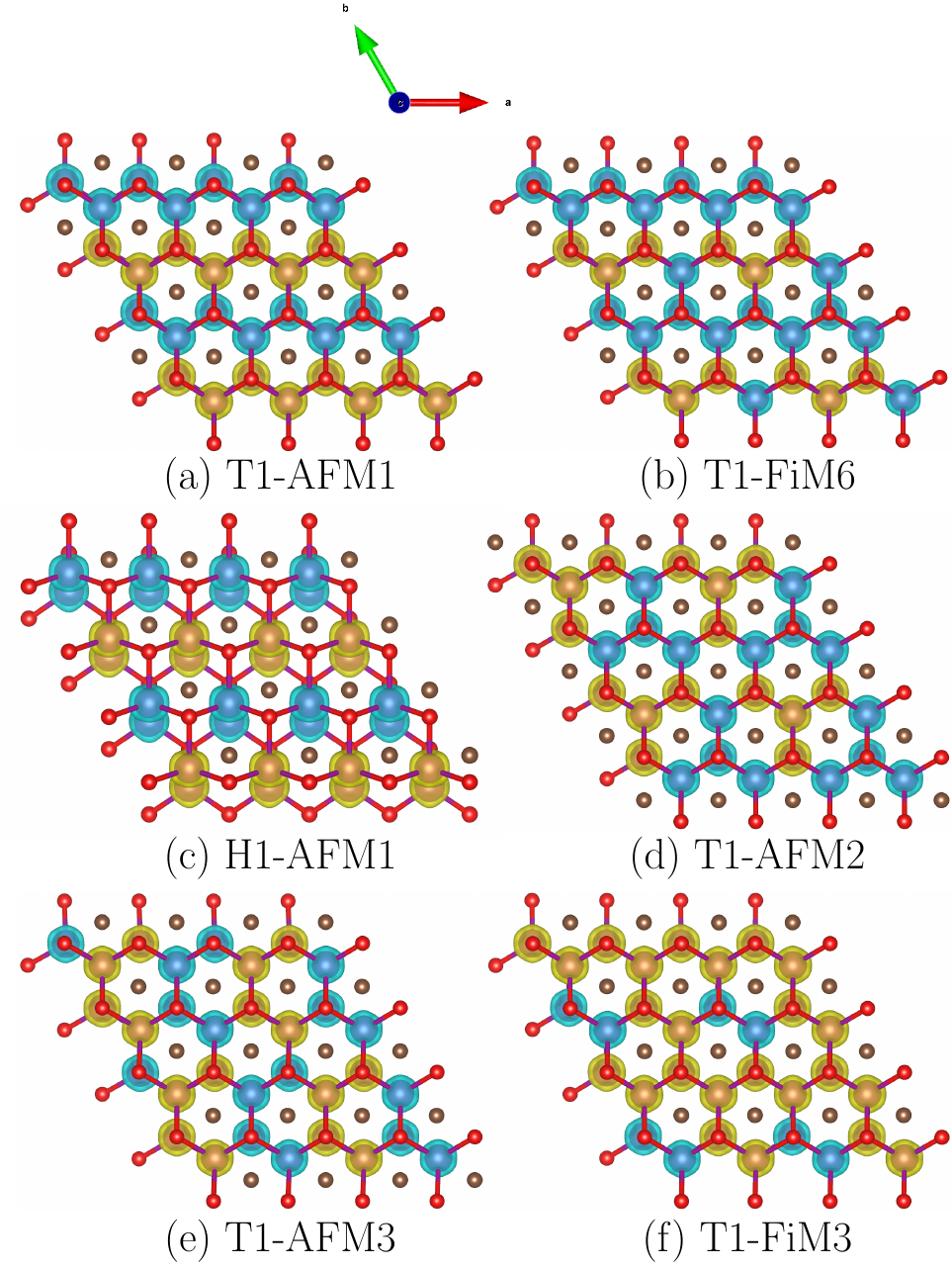}
  \caption{Top six most energetically favorable spin solutions for the \ce{Mn2CO2} MXene. Different colors correspond to spin density - the residual spin-up (yellow) or spin-down (blue) on each transition metal. The color code for individual atoms is identical to Figure \ref{geom_isomers}.}
  \label{conformers}
\end{figure}

To learn more about magnetism in the Mn$_2$CO$_2$ system, we have created new spin solutions in $2\times2$ supercells (Mn$_8$C$_4$O$_8$) corresponding to both T1 and H1 structures, namely seven additional antiferromagnetic (AFM) and eight additional ferrimagnetic (FiM) conformers per geometry.  
The visualizations and labels (combining geometry from Figure \ref{geom_isomers} and particular magnetic states, e.g., T1-AFM1) of all such spin conformers can be seen in Electronic Supplementary Information (ESI, Figures S1 -- S4). 
With the inclusion of standard unit cells (simple $1\times1$ magnetic motifs, denoted as T1/H1-AFM0 and T1/H1-FM0), we obtained 34 structures in total, reducing finally to 28, because some spin states were identical to others due to symmetry. 
The first approach was to investigate the spin solutions with the GGA PBE density functional (ESI Table S1, the PBE ground-state was H1-AFM6 solution). However, the metallic nature of some spin states and the behavior of their band structures was finally classified as an artifact of GGA PBE band underestimation (see more details in ESI), and more reasonable meta-GGA SCAN density functional was used. 
The use of meta-GGA functional for $2\times2$ supercells with 20 atoms in the computational cell was a reasonable choice because meta-GGA SCAN functionals often provide the band gaps that are in good agreement with DFT+U or hybrid functionals with only a marginal increase in computational time comparing to GGA density functionals (see also discussion below.) 
The final energetics and band gaps provided by SCAN density functional are collected in Table \ref{SCAN_table}, and the top six conformers are visualized in Figure \ref{conformers}. Note that the lattice constant of all spin solutions corresponds to the ground state. Table \ref{SCAN_table} also includes results based on additional hybrid density functional HSE06 calculations. 

\begin{table}[ht]
    \centering
    \caption{Relative energies (R.E.) per unit ($1\times1$) cell at the level of meta-GGA SCAN density functional and hybrid functional HSE06 for T1 and H1 \ce{Mn2CO2} conformers and all 28 calculated spin solutions. Each column of relative energies is followed by an indirect ($E_{\mathrm{g}}^{\mathrm{indir}}$) and direct ($E_{\mathrm{g}}^{\mathrm{dir}}$) electronic band gap at the corresponding level of theory. The lattice constant is fixed at $a = 2.89 \ \ag$ for all solutions. For spin-polarized FM and FiM solutions, the lower value of the band gap is reported. All energies are presented in eV.}
    \begin{tabular}{cccccc}
    \hlineB{1}
        ~ & \multicolumn{2}{c}{SCAN} & \multicolumn{2}{c}{HSE06} \\
        Phase & R.E. & $E_{\mathrm{g}}^{\mathrm{indir}} \ (E_{\mathrm{g}}^{\mathrm{dir}})$ & R.E. & $E_{\mathrm{g}}^{\mathrm{indir}} \ (E_{\mathrm{g}}^{\mathrm{dir}})$\\
    \hlineB{1}
        T1-AFM1	&0.000	& 0.73 (0.92) & 0.000 & 1.53 (1.61) \\
        T1-FiM6	&0.017	& 0.73 (0.87) & 0.051 & 0.86 (1.40) \\
        H1-AFM1	&0.018	& 0.26 (0.41) & 0.072 & 0.68 (1.03) \\
        T1-AFM2	&0.023	& 0.86 (1.13) & 0.043 & 1.48 (1.95) \\
        T1-AFM3	&0.024	& 0.64 (1.03) & 0.096 & 0.76 (1.45) \\
        T1-FiM3	&0.031	& 0.66 (0.66) & 0.078 & 0.80 (1.21) \\
        T1-AFM4	&0.037	& 0.58 (0.73) & 0.182 & 0.49 (1.12) \\
        T1-FiM4	&0.037	& 0.53 (0.70) & 0.140 & 0.63 (1.22) \\
        T1-AFM5	&0.041	& 0.74 (0.92) & 0.123 & 0.81 (1.46) \\
        H1-FiM4	&0.042	& 0.28 (0.48) & 0.109 & 0.51 (0.99) \\
        H1-FiM3	&0.043	& 0.28 (0.51) & 0.103 & 0.33 (0.76) \\
        T1-FiM5	&0.046	& 0.64 (0.76) & 0.104 & 0.87 (1.31) \\
        H1-AFM3	&0.047	& 0.40 (0.79) & 0.124 & 0.77 (1.37) \\
        H1-AFM2	&0.049	& 0.38 (0.78) & 0.131 & 0.51 (0.89) \\
        T1-FiM1	&0.049	& 0.37 (0.52) & 0.168 & 0.50 (1.08) \\
        H1-FiM1	&0.053	& 0.37 (0.65) & 0.130 & 0.66 (0.96) \\
        H1-FiM5	&0.056	& 0.52 (0.75) & 0.144 & 0.47 (0.96) \\
        T1-FiM2	&0.056	& 0.54 (0.54) & 0.110 & 0.91 (1.17) \\
        H1-AFM6	&0.057	& 0.58 (1.03) & 0.172 & 0.88 (1.60) \\
        H1-FiM8	&0.057	& 0.41 (0.88) & 0.153 & 0.74 (1.30) \\
        T1-FiM7	&0.057	& 0.83 (0.95) & 0.115 & 0.98 (1.52) \\
        H1-FiM2	&0.060	& 0.48 (0.55) & 0.131 & 0.51 (0.89) \\
        H1-FiM7	&0.067	& 0.50 (0.90) & 0.173 & 0.74 (1.22) \\
        H1-FM0	&0.069	& 0.45 (0.45) & 0.126 & 0.41 (0.85) \\
        H1-AFM7	&0.069	& 0.67 (1.07) & 0.187 & 1.05 (1.60) \\
        T1-FM0	&0.079	& 0.47 (0.47) & 0.125 & 0.98 (1.11) \\
        H1-AFM0	&0.082	& 0.70 (1.06) & 0.203 & 0.98 (1.62) \\
        T1-AFM0	&0.085	& 1.25 (1.11) & 0.139 & 1.57 (2.13) \\
    \hlineB{1}
    \end{tabular}
    \label{SCAN_table}
\end{table}

The ground state corresponds to the T1-AFM1 conformer (Table \ref{SCAN_table}; in contrast to the PBE results of Table S1): trigonal geometrical structure with terminal oxygen atoms in hollow sites and spins localized on Mn atoms in spin-up/spin-down zigzag lines. 
T1-AFM1 solution is the ground state at both SCAN and HSE06 levels of theory. The order of other spin states is slightly altered, but the top six spin solutions remain on top, and the energy differences between individual solutions increased significantly when HSE06 density functional was used. Speculatively, a higher, more correlated level of theory could make T1-AFM1 even more pronounced ground state.
These results agree well with recent theoretical studies showing ground-state spin configuration equivalent to our T1-AFM1.\cite{He2016,Gao2022} 
On the other hand, predicted ferromagnetic ground-state and conducting antiferromagnetic states in other studies\cite {Kozak2023} can be the result of a limited search at the PBE level with different geometries. 
Interestingly, our original $1\times1$ unit cell conformers (T1-FM0, T1-AFM0, and H1-AFM0) are the lowest ones in our Table \ref{SCAN_table}, implying that the investigation of only unit cell of \ce{Mn2CO2} is leading to biased knowledge: ferromagnetic solution (H1-FM0) would be the false ground state with much smaller band gap than T1-AFM1. 
Allowing energy-lowering using the right spin ordering (supercell motifs) is crucial for studying the \ce{Mn2CO2} system, similarly to, e.g., antiferromagnetic mackinawite FeS.\cite{Kolos2020} Such a trend of the antiferromagnetic state overtaking the ferromagnetic phase does not apply only to the oxygen-terminated MXene. As has been recently shown in multiple studies, even the precursor MAX phase \ce{Mn2GaC} has an AFM ground-state when more complex spin motifs are taken into account\cite{Ingason2014, Dahlqvist2020, Novoselova2018, Dahlqvist2016}. The same result holds even for the clean \ce{Mn2C} sheet, although both the MAX phase and \ce{Mn2C} are conductors.\cite{Zhou2017, Zhang2020, Hu2016} The semiconducting behavior is therefore only achieved by oxygen termination. The antiferromagnetic ground state can be then also seen in other non-terminated MXenes, such as two-, three-, or four-layer Ti$_n$C$_{n-1}$ sheets.\cite{Lv2020, Gorkan2022, GarcaRomeral2023, Venosova2024}

As can be seen from Table \ref{SCAN_table}, all Mn$_2$CO$_2$ conformers are semiconducting. 
We have also used the total energies to estimate the probabilities of Mn$_2$CO$_2$ being in a given state, which were evaluated for different temperatures and can be seen in the last three columns of Table S2 of ESI: the simple Boltzmann factor prefers only the first conformer to appear at room temperature. 

To confirm the stability of the conformers, more realistic energy differences, and their energy ordering, we have investigated the top five conformers (T1-AFM1, T1-FiM6, H1-AFM1, T1-AFM2, and T1-AFM3) using ab initio molecular dynamics simulations at 400 K. 
Results showing the mean energies with highlighted standard deviation can be seen in ESI. 
From these results, we can conclude that the molecular dynamics simulations indicate no phase change under 400 K. 
I.e., the T1-AFM1 is confirmed as the ground-state conformer. 
This is in line with experimentally prepared parent \ce{Mn2GaC} MAX phase holding AFM ordering at higher temperatures with Néel temperature of 507 K.\cite{Novoselova2018} 
We have also calculated the phonon dispersion spectra for some of these conformers to further confirm their stability (see ESI for no negative frequencies appearance). 
Due to small total energy differences (ca 0.017~eV), Mn$_2$CO$_2$ MXene is, therefore, a promising material with its potential AFM-FiM switching properties: The energetically lowest stable antiferromagnetic conformer is the T1-AFM1 conformer (the electronic ground state), and the lowest stable ferrimagnetic conformer is the T1-FiM6 conformer (the electronic first excited state), both with the same geometry (T1) differing just in the magnetic motif (cf. Figure \ref{conformers}).

Further, the biaxial strain on the five energetically most favorable configurations was simulated to see, how the indirect and direct electronic band gap changes. 
The results of these strain calculations can be seen in ESI Figure S10 -- we have also marked the $\Gamma$-point gap to see where the direct electronic band gap moves to a different point in k-space.  
From the results, we can see that the T1-AFM2 configuration should undergo a transition to direct material under $-4\%$ compressive strain. 
Interesting is the result for configuration H1-AFM1 which exhibits conducting behavior under strain as the indirect gap vanishes. 
This behavior is similar to the Ti$_2$CO$_2$ MXene, for which the direct band gap also decreases to zero under compressive -4\% biaxial strain.\cite{Ding2020, Li2020}

\subsection{Density functional theory analysis of the ground state configuration} \label{dft-gs}

For a further in-depth study of Mn$_2$CO$_2$, we have decided only to focus on the ground-state configuration T1-AFM1. 
We have made this decision because the energy difference between the top two configurations (0.017 eV) is the largest of all tested configurations, with the rest being mostly in the range of 3 - 8 meV and the smallest being only 0.1 meV. 
This decision is further confirmed by our molecular dynamics results discussed above. 
Also, Table \ref{SCAN_table} shows that the individual energy levels are spaced further apart when a hybrid HSE06 density functional is used. The energy difference between the ground state and the second spin solution is then increased more than two-fold. 
Lastly, the reason for focusing only on the ground state was that the following GW+BSE calculations are highly computationally demanding.

\begin{figure}[h]
    \centering
    \includegraphics[height=10cm]{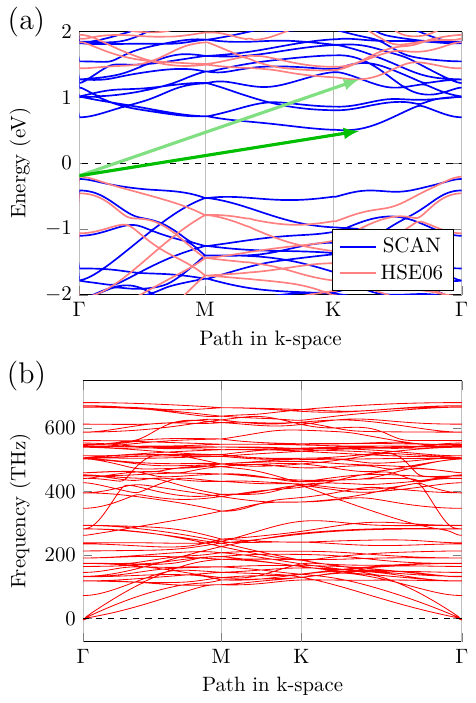}
    \caption{T1-AFM1 band structure at the meta-GGA SCAN density functional and hybrid HSE06 levels. The Fermi energy is set to zero. The green arrows in the band structures signify the HSE06 and SCAN indirect electronic band gap. (b) Phonon dispersion spectra of T1-AFM1 show no signs of instability.}
    \label{t1afm1_results}
\end{figure}

We present the SCAN and additional HSE06 band structures and phonon dispersion spectra of the ground-state conformer T1-AFM1 of Mn$_2$CO$_2$ in Figure \ref{t1afm1_results}. 
The bands corresponding to a different spin are not distinguishable in the spin-polarized band structure near the Fermi level due to the antiferromagnetic nature of the material.
We can observe quite flat bands around the Fermi level and an indirect band gap with the smallest energy transition being from the $\Gamma$ point to near the K point. 
Based on the recommended values of the Hubbard correction U for the Mn-based MXenes from previous studies,\cite{He2016, Hu2016, Dahlqvist2020} we conclude from band structures (Figure S11) and band gaps (Table S3) that PBE+U (U = 3 eV) corresponds well with a non-empiric SCAN approach without any U correction. 
This agreement can be further seen in the partial density of states (PDOS; Figure S12) where the oxygen contribution to the valence states using SCAN functional corresponds well with the PBE+U(3 eV), and treatment of 3$d$ orbitals from SCAN seems consistent with PBE+U(3 eV), while over-performed by PBE and underestimated by HSE06. 
Moreover, magnetic moments provided by SCAN density functional ($\sim$2.7$\mu_B$ per Mn atom; Table S3) are smaller than ones from PBE+U(3 eV) or HSE06 functionals ($\sim$3$\mu_B$), which is right trend compared to experimental measurements on \ce{Mn2GaC} MAX precursor. 
SCAN density functional, therefore, seems suitable for the DFT description of the Mn-based MXenes.
Based on the Bader charge analysis we can conclude that the manganese atoms lose approximately 1.7 $e$ (charge of an electron) by bonding to the \ce{Mn2CO2}. On the other hand, each terminal oxygen has acquired a charge of 1 $e$ and a similar situation is for carbon atoms which gained approximately 1.4 $e$ of charge.

\begin{figure}[h]
\centering
    \hspace{-1cm}
    \includegraphics[width=.6\linewidth]{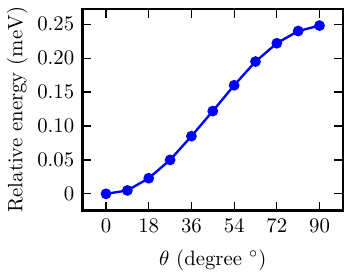}
    \caption{The angular dependence of the unit cell total energy for T1-AFM1 conformation of the \ce{Mn2CO2} MXene. The $\theta$ is the angle between the magnetic moment and the normal to the monolayer plane.} 
    \label{mae_path}
\end{figure}

Because magnetism is the dominant feature of Mn$_2$CO$_2$ material, we have investigated the magnetic anisotropy energy (MAE) for the T1-AFM1 conformation, which determines the orientation of the magnetization at low temperatures. 
We performed non-collinear total energy calculations with the inclusion of the spin-orbit coupling for a path between two spin orientations, $\theta = 90^{\circ}$ and $\theta = 0^{\circ}$, where the angles denote the magnetization in-plane and perpendicular to the monolayer MXene, respectively. The MAE is then calculated by
\begin{equation}\label{mae}
    \mathrm{MAE} = E(\theta=90^{\circ}) - E(\theta=0^{\circ}).
\end{equation}
Having the easy axis set to the $z$-direction ($\theta = 0^{\circ}$), we have calculated the MAE of the \ce{Mn2CO2} MXene to be 0.25 meV per unit cell. The full dependence of the total energy on the angle $\theta$ is shown in Fig \ref{mae_path}. Therefore, the preferred direction of magnetization is along the easy axis. Interestingly, it has been shown that the pure monolayer \ce{Mn2C} without terminations prefers the in-plane magnetization (with the MAE of -0.09 meV per unit cell),\cite{Hu2016} similarly as the precursor MAX phase \ce{Mn2GaC}.\cite{Dahlqvist2020} Our result corresponds quite well with a previous study where the MAE for \ce{Mn2CO2} has been calculated to be 0.18 meV per unit cell.\cite{He2016} MAE of 2D \ce{Mn2CO2} is larger than that of pure metals, such as Fe and Ni, as for them the reported values of MAE per unit cell are 1.4 $\mu$eV and 2.7 $\mu$eV, respectively.\cite{Daalderop1990} These advantages render the MXene sheet a very promising candidate for AFM spintronic nanodevices.\cite{Hu2016, He2016} It is obvious that the lowered dimensionality of 2D materials leads to an increase in MAE as a similar effect can be observed in various families of 2D materials.\cite{Jiang2021}

\subsection{Many-body calculations} \label{many-body}

The acquisition of credible GW, and consequently BSE results is highly dependent on the proper convergence of the simulation parameters.\cite{Dubecky2020, Dubecky2023, Kolos2022} 
These parameters include the energy cutoff for the response function used in the GW calculation, $E_{\mathrm{cut}}^{\mathrm{GW}}$, the number of bands, $N_{\mathrm{B}}$, the number of frequency grid points, $N_{\omega}$, the size of the k-point grid in the $xy$-direction or the inter-sheet distance in the $z$-direction. 
It has been recently shown\cite{Dubecky2023, Fanta2023} how the inter-sheet distance can deform the lowest unoccupied orbitals and therefore give untrustworthy many-body results.
We have therefore made a partial decomposition of bands to see how prominent the orbitals were in the $z$-direction. As it turned out, orbitals in our system were mostly localized on transition metals (see ESI Figure S13). During our convergence calculations, we observed, that Mn$_2$CO$_2$ was highly sensitive to computational parameters used in the GW step. 
Quasiparticle (QP) band gap, and consequently optical gap, could vary significantly (in orders of electronvolts) if less strict parameters were used. Therefore, a careful approach to settings was essential.

The convergence of optical absorption spectra and optical gap $\Delta_{\mathrm{opt}}^{\mathrm{BSE}}$ with respect to the number of considered GW bands that are used in the solution of the BSE is reported in ESI Figure S14. As can be seen, 16 occupied and 20 virtual bands used in the BSE step were sufficient for reliable absorption spectra for photon energies up to 5 eV. Therefore for all subsequent computations, these bands were used. Following in ESI Figure S15 are convergences of optical absorption spectra (for $A_{xx}=A_{yy}$) with respect to several computational parameters and the convergence characteristics of QP band gap ($\Delta_{\mathrm{f}}^{\mathrm{GW}}$) and optical gap ($\Delta_{\mathrm{opt}}^{\mathrm{BSE}}$) on those same parameters. The benchmarking approach for Mn$_2$CO$_2$ was to use the number of bands $N_{\mathrm{B}}=3024$, GW energy cutoff $E_{\mathrm{cut}}^{\mathrm{GW}}=200$ eV, $6\times6\times1$ k-point grid, the height of the computational cell of $L_z=20 \ \ag$ and the number of frequency-dependent grid points $N_{\omega}=128$, where the respective parameter was gradually changed in each convergence study. These parameters were used to avoid extremely demanding GW calculations. The obtained spectra show that in almost all calculations we could observe an emergence of a new peak accompanied by a shift of $\Delta_{\mathrm{opt}}^{\mathrm{BSE}}$. The only exception was the convergence concerning the k-point grid, where we did not observe any significant changes in both $\Delta_{\mathrm{f}}^{\mathrm{GW}}$ and $\Delta_{\mathrm{opt}}^{\mathrm{BSE}}$. Usually $\Delta_{\mathrm{f}}^{\mathrm{GW}}$ is quite sensitive to change in k-point grid.\cite{Dubecky2020,Dubecky2023, Qiu2013}

Both $\Delta_{\mathrm{f}}^{\mathrm{GW}}$ and $\Delta_{\mathrm{opt}}^{\mathrm{BSE}}$ converged sufficiently in most cases with the direct fundamental gap $\Delta_{\mathrm{f}}^{\mathrm{GW}}$ changing in the range of $\Delta_{\mathrm{f}}^{\mathrm{GW}}=1.92-2.20$ eV and optical gap in the range $\Delta_{\mathrm{opt}}^{\mathrm{BSE}}=1.44-1.65$ eV. The most problematic computational parameter for the estimation of $\Delta_{\mathrm{f}}^{\mathrm{GW}}$ is the height of the computational cell $L_z$, where the change in the gap across calculations was 0.21 eV. As the fundamental gap $\Delta_{\mathrm{f}}^{\mathrm{GW}}$ was not fully converged, we fitted the values with the linear fit\cite{Dubecky2023, Berseneva2013, Choi2015, Karlicky2018}
\begin{equation}
    \Delta_{\mathrm{f}}^{\mathrm{GW}}\left(\frac{1}{L_z}\right) = C\frac{1}{L_z} + \Delta_{\mathrm{f}}^{\mathrm{GW}}(0),
\end{equation}
where $C$ and $\Delta_{\mathrm{f}}^{\mathrm{GW}}(0)$ are fitting parameters. The extrapolation to the zero $1/L_z$ limit yielded the direct gap $\Delta_{\mathrm{f}}^{\mathrm{GW}}$ of 2.52 eV, therefore for the subsequent final estimation of the fundamental band gap a corresponding \textit{a posteriori} rigid correction was used, by taking $\Delta_{\mathrm{f}}^{\mathrm{GW}}(0)-\Delta_{\mathrm{f}}^{\mathrm{GW}}(1/L_z)=0.40$ eV, where $1/L_z=0.04 \ \ag^{-1}$ was used. Similarly we fitted the values of the optical gap $\Delta_{\mathrm{opt}}^{\mathrm{BSE}}$ with the linear fit and the extrapolation to the zero $1/L_z$ limit yielded $\Delta_{\mathrm{opt}}^{\mathrm{BSE}}=1.43$ eV and therefore \textit{a posteriori} rigid correction of $\Delta_{\mathrm{opt}}^{\mathrm{BSE}}(0)-\Delta_{\mathrm{opt}}^{\mathrm{BSE}}(1/L_z)=-0.11$ eV was used in the final estimation of the optical gap.

Following from the presented convergence results we can see that for the final production estimate the settings of $N_{\mathrm{B}}=5904$, $E_{\mathrm{cut}}^{\mathrm{GW}}=250$ eV, $6\times6\times1$ k-point grid, $L_z=25 \ \ag$ and $N_{\omega}=216$ are sufficient for precise results. 
With this calculation, we can arrive at the final production estimate (with the correction to $L_z$) of the direct (d) and indirect (i) GW quasiparticle band gap for monolayer Mn$_2$CO$_2$ MXene of
\begin{equation}
\begin{aligned}
    & \Delta_{\mathrm{f}}^{\mathrm{GW, d}} = 2.39 \ \mathrm{eV}\\
    & \Delta_{\mathrm{f}}^{\mathrm{GW, i}} = 2.05 \ \mathrm{eV},
\end{aligned}
\end{equation}
and the estimate of the BSE optical gap (again with the corresponding correction to $L_z$)
\begin{equation}
    \Delta_{\mathrm{opt}}^{\mathrm{BSE}} = 1.32 \ \mathrm{eV}.
\end{equation}
From the direct QP gap and optical gap the first exciton binding energy consequently amounts to $E_{\mathrm{b}}^{\mathrm{BSE}}=1.07$ eV which is an unusually large amount as many 2D semiconductors show the linear scaling between the $\Delta_{\mathrm{f}}$ and $E_{\mathrm{b}}$, $E_{\mathrm{b}}\approx\Delta_{\mathrm{f}}/4$, which is independent on the lattice configuration, bonding characteristics or topological properties.\cite{Jiang2017}

\begin{figure}[h]
\centering
\includegraphics[height=6cm]{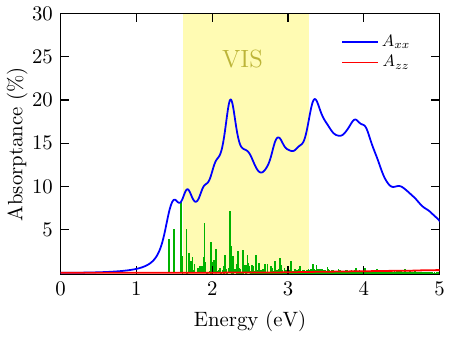}
\caption{Optical absorption spectrum ($A_{xx}=A_{yy}$) for monolayer T1-AFM1 at the level of G$_0$W$_0$@SCAN+BSE and corresponding oscillator strengths. The yellow region signifies the visible electromagnetic spectrum. The results were obtained with $6\times6\times1$ k-point grid, $E_{\mathrm{cut}}^{GW}=250$ eV, $L_z=25 \ \ag$, $N_{\mathrm{B}}=5904$, $N_{\omega}=216$ and 16 occupied and 20 unoccupied bands used in the final BSE step.} 
\label{spectra_a1afm1}
\end{figure}

The final optical absorption spectrum in planes parallel and perpendicular to the MXene sheet is reported in Figure \ref{spectra_a1afm1}.
The visualized quantity is the absorptance $A(E) = 1-\mathrm{exp}[-\epsilon_2 E L_z/{\hbar}c]$,\cite{Ketolainen2020} where $E$ is the energy of incoming photon, $\epsilon_2$ is the imaginary part of dielectric function, $\hbar$ is reduced Planck's constant, and $c$ is the speed of light. 
Absorptance in the visible spectrum of photon energy (1.63-3.26 eV) is from Figure \ref{spectra_a1afm1} estimated to amount to $A\approx 10-20\%$ which shows that Mn$_2$CO$_2$ could be a very good solar light absorber. Similar efficiency is observed for the absorptance of near UV radiation (up to 4.13 eV). Moreover, the visible radiation absorption fully covers the whole range which is a significant improvement over different MXenes (namely, Cr- or Sc-based) for which only a small region of the visible spectrum is absorbed and which would make the potential photoabsorption devices less effective.\cite{Ketolainen2022} 
Mn$_2$CO$_2$ MXene is, therefore, a very promising material with its potential AFM-FiM switching properties (due to small total energy differences) and exceptional absorption efficiency.

\begin{figure}[h]
 \centering
 \includegraphics[height=3cm]{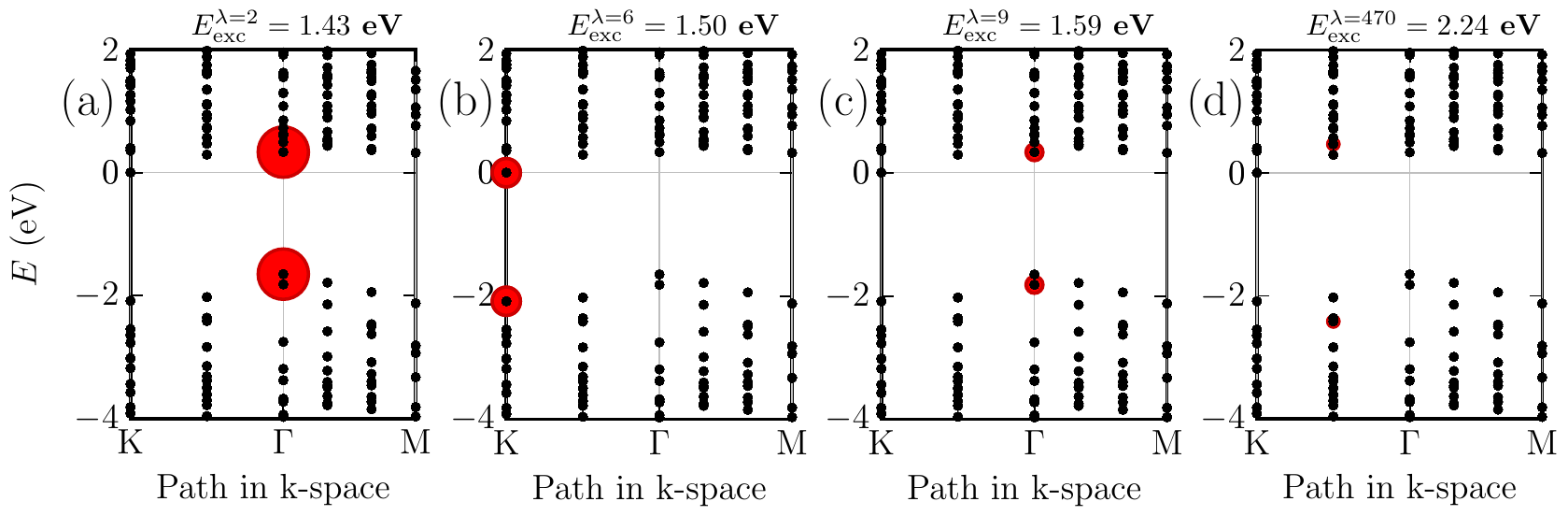}
 \caption{Quasi-particle (GW) band structure of T1-AFM1 Mn$_2$CO$_2$ MXene and all $|A_{cv\mathbf{k}}^{\lambda}|$ coefficients for several bright excitons. The radii of colored circles represent coefficients. The Fermi energy is set to zero.}
 \label{fatbands}
\end{figure}

Finally, we have analyzed (Figure \ref{fatbands}) the excitonic wave functions (Equation \ref{exciton_wf}) of four bright excitons important in the optical spectra (Figure \ref{spectra_a1afm1}). 
The coefficients $|A_{cv\mathbf{k}}^{\lambda}|$ correspond to the contribution of a given electron-hole pair at a certain k-point and band to the exciton wave function. Interestingly, for all excitons in Figure \ref{fatbands} mostly only one electron-hole pair dominates the excitonic state. The first optical transition ($E_{\mathrm{exc}}^{\lambda=2}$, as the exciton with the energy $E_{\mathrm{exc}}^{\lambda=1}$ is dark) is dominated by the electron-hole pair in the highest occupied band and the lowest unoccupied band in the $\Gamma$ point while the second transition ($E_{\mathrm{exc}}^{\lambda=6}$) is dominated by the pair in the K point. This corresponds with the indirect character of Mn$_2$CO$_2$ and flat bands around the Fermi level.

\section{Conclusions}
We have systematically investigated geometrical and spin conformations of oxygen-terminated \ce{Mn2C} MXene and performed subsequent many-body calculations to obtain reliable electronic and optical properties. 
The Mn-based MXenes are perspective because of the broad palette of its magnetic phases, and the parent MAX phase was already prepared (\ce{Mn2GaC}).\cite{Ingason2014, Novoselova2018, Thorsteinsson2023}   
Typical calculations from basic (GGA) density functional theory (DFT) indicated the conducting behavior of \ce{Mn2CO2},\cite{Champagne2021, Kozak2023} while recent studies predicted it also semiconducting.\cite{Ketolainen2022,He2016, Gao2022, Chen2021, Zhou2017} 
To reconcile such uncertainties, we systematically generated a large set of \ce{Mn2CO2} magnetic solutions using meta-GGA density functional SCAN and hybrid density functional HSE06, proving that all conformers embody the semiconducting behavior (while band gap varied). 
Different magnetic states and geometrical conformations were energetically very close, but molecular dynamics simulations have shown us that even at room temperature there should be no spontaneous phase switching. 
The ground-state \ce{Mn2CO2} conformation is trigonal geometry with terminal oxygen atoms in hollow sites and antiferromagnetic (AFM) spin alignment of Mn atoms in spin-up/spin-down zigzag lines. 
The AFM ground state is, therefore, consistent with the experimentally prepared parent MAX phase \ce{Mn2GaC} (AFM ordering up to Néel temperature of 507 K)\cite{Novoselova2018} and even with the clean non-terminated MXene sheet \ce{Mn2C}.
Moreover, this configuration exhibits a strong preference for the magnetization direction along the $z$-axis with magnetic anisotropy energy (MAE) 0.25 meV. 
Ignoring local magnetic motifs (unit cell use) led to the wrong ferromagnetic ground state with different electronic properties and allowing the energy lowering by supercell motifs was fundamental for \ce{Mn2CO2} material.

For the ground state, we performed a series of subsequent many-body GW and Bethe-Salpeter equation (BSE) calculations. 
We showed the \ce{Mn2CO2} as an indirect semiconductor with a fundamental gap of 2.1 eV (and a direct gap of 2.4 eV) and the first bright optical transition at 1.3 eV. 
The binding energy of the first exciton (1.1 eV) is very high and is almost half of the direct gap. 
Unlike other MXenes, \ce{Mn2CO2} absorbs the whole visible light range and near UV range efficiently (between 10 - 20\%). 
We can therefore classify the \ce{Mn2CO2} MXene as a semiconducting antiferromagnet and efficient visible light absorber.


\section*{Conflicts of interest}
There are no conflicts to declare.

\section*{Acknowledgements}
This work was supported by the Czech Science Foundation (21-28709S) and the University of Ostrava (SGS06/PrF/2023, SGS04/PrF/2024). The calculations were performed at IT4Innovations National Supercomputing Center (e-INFRA CZ, ID:90140).



\balance


\bibliography{rsc} 
\bibliographystyle{rsc} 

\end{document}